\documentclass[12pt]{article}
\usepackage{amsmath}
\usepackage{graphicx}
\usepackage{amssymb}
\begin{document}

\renewcommand{\title}[1]{%
    \bigskip%
    \begin{center}%
    \Large\bf #1%
    \end{center}%
    \vskip .2in}

\renewcommand{\author}[1]{%
   {\begin{center}
    #1
    \end{center}}}
\newcommand{\address}[1]{\vspace{-1.7em}\vspace{0pt}
    {\begin{center}
    \it #1
    \end{center}}}
 
\renewcommand{\title}[1]{%
    \bigskip%
    \begin{center}%
    \Large\bf #1%
    \end{center}%
    \vskip .2in}

{\title{\bf{ Covariant fracton gauge theory }}}

\author
{  
Sk. Moinuddin   $\,^{\rm a, b}$, 
Pradip Mukherjee  $\,^{\rm a,c}$}

\address{$^{\rm a}$Department of Physics, Barasat Government College,Barasat, India}
\address{$^{\rm b}$\tt
dantary95@gmail.com  }
\address{$^{\rm c}$\tt mukhpradip@gmail.com}

\vskip1.5cm
 
\abstract{ 
  A paradigm shift of the fracton 
physics research is established by  providing  a covariant formulation of the action. For the first time, fracton gauge symmetries are connected with the global symmetries of the free dynamics. A Galileon  doublet are assumed to construct the scalar charged 
  matter field. The corresponding action is shown to be the repository of all the symmetries of the system.
  The symmetric tensor gauge field introduced during  localization of the gauge symmetries is the empirical
  tensor gauge field known today. All the past paradoxes are shown to dispel by the 
use 
of Galileon scalar field doublet in place of the usual scalar field doublet as the basic building blocks of 
matter action.
}
  
 \vskip1.5cm

\begin{enumerate}

\item
  {\bf{ Introduction:}}

  In this letter we propose a paradigm shift in the ongoing research in fracton gauge theories. Fracton is a quasi particle following from certain exactly solvable three dimensional spin and Majorana models \cite{cham} \cite{haa}  \cite{vij}. The distinctive features of the fractons from all other quasi particles is their very low
(practically vanishing)  mobility. Experimental signature of the fracton like-behavior  is being reported now \cite{cham}\cite{pret3}. The immobility of the fractons are considered to be useful   for quantum information storage \cite{ Haa} \cite{ ser}. To understand the vanishing mobility one has to assume
that in addition to the conserved charge
the dipole moment is also to be conserved.
What is the cause  of the new symmetry 
was unknown to the community. Pretko \cite{pret} constructed the second symmetry by a procedure which was called `fracton gauge theory' . However in the following we will see that the methodology of \cite{pret} can hardly be recognized as a gauge theory. The reason is it was not constructed using the basic tenets of gauge theory formulation \cite{d}.  Actually, gauge symmetry follows from the global internal symmetry of the matter action. Thus the paradigm shift is draw the attention of the community to emphasize the matter sector. The present letter will amply justify it.

 \hskip 1 cm  Another point of departure is the lack of covariance in the existing formulations . In this letter we provide a covariant formulation of the fracton gauge theory which successfully explains the basis of the different symmetries of the fracton phase of glassy systems . Note that this is the first covariant formulation in the literature . The voluminous information about the system were accumulated by many authors \cite{pret} \cite{and} paved the emergence of our theory.  All the basic elements of experimental and theoretical side for a covariant analysis were ready.  

\hskip 1 cm  In this letter we will show that the fracton gauge theory  can be obtained   from a covariant theory of doublet of Galileon scalar fields. There are five Galileon actions
  $\mathcal{L}_1 $to
  $\mathcal{L}_5 $. Of them 
  $\mathcal{L}_1 = \pi$ is trivial and 
  $\mathcal{L}_2 = \partial_\mu\pi \partial^\mu\pi$ is the usual Lagrangian . $\mathcal{L}_3$ , $\mathcal{L}_4$  and $\mathcal{L}_5$ exhibit all the Galileon features.Despite being higher derivative theories, Galileons are free of Ostogradsky-type ghosts since they have no more than second order equations of motion. The Galileon  field is a scalar under Poincare transformations and  has a unique symmetry under the shift,
 
     \begin{equation} 
\pi \to \pi + a +b_{\mu} x^\mu \label{0}
\end{equation}

  where a and $b_{\mu}$  are  real
     numbers. Note that we some times indicate the first term $a$ as a translation and second part $ b_{\mu} x^\mu$  as shift while some time we will refer the whole transformation as shift. It will understood from the context, which meaning is implied.      
  
\hskip 1 cm   We will take up to the third Galileon in our model. This  apparently abstract exercise will reveal a lot of information regarding the fracton phases of matter, discovered in connection of the glassy material \cite{cham}. The brilliant hypothesis of Pretko \cite{pret}   and an intense effort in that line, constructed the essential features of a tensor gauge theory appropriate for fracton phenomenology. Other approaches also indicated to a symmetric tensor gauge theory. However a basic understanding of the unusual symmetry of the fracton system is still 	
    eluding us.  The essential physics behind the additional symmetry which is exposed in the experimental observation still could not be explained from some basic theory. Our results may really be useful in this connection.

    To place our results in the proper perspective  we begin with the salient features of existing `fracton gauge theory'.

\item {\bf{ The  existing fracton gauge theory - - our observations  \cite{pret1}\cite{nand}}}

  First of all the procedure deviates from the standard gauge theory formulation in many respects. It is well
  known that in a gauge invariant theory of matter (interacting with gauge field)
  there are two parts, one part describing
  the matter sector and  another the gauge dynamics. In addition there is one or more
  terms containing a coupling of the matter
  field with the gauge field. Not only this ,
  the whole theory must be gauge and space-time transformations invariant , other wise the gauge symmetry  loses its meaning.
  
\hskip .5cm  Secondly the matter sector is completely absent in the formulation. Actually, in a
gauge theory, the matter field must carry a representation of the global symmetry group. Here there is no emphasis on this .
In fact the time parameter is absent in the
consideration of co-variance. The result in
the theory may loses time translation invariance.

\hskip .5cm Having listing the short comings, we must emphasize the positive points of \cite{pret1}\cite{nand}. These are the explanation of the peculiar low mobility of the fracton by the two conserved currents corresponding to the two symmetries. The reason behind the
success of the existing  theory as a phenomenology owes to it. But a full fledged fracton gauge theory must explain why these symmetries  appear here along with the emergence of the co-variant dynamics. We have provided a consistent theory of the fracton phases which is both Poincare and gauge symmetric.


\item
{\bf{ Galileon based Fracton gauge theory :}}

 (i) {\bf{The fracton phase of matter:}} 
 
The fracton physics comprises  a class of new phenomena where the excitations were characterized by unusually small mobility \cite{haa}  \cite{vij}. Obviously this indicates some unknown symmetry opposing the motion of the excitation . It was subsequently identified as the conserved dipole moment \cite{pret}, in addition to the usual charge conservation. Mathematically,
\begin{eqnarray}
 Q = \int \rho d^D x  =  {\rm{ constant }} ; J^k \hskip .4cm = \int \rho x^k d^D x
 \label{20}
\end {eqnarray}

The charge conservation theorem is a property of the electromagnetic theory. Thus any matter theory must have a conserved charge to interact with the electromagnetic field. Note that the first conservation law gives the second a coordinate independent meaning. These  hierarchy of the conservation laws is to be noted carefully.

 \hskip 1 cm        While it is known that the usual global phase rotation symmetry  leads to the charge conservation, the symmetry from which the conservation of the dipole moment  emerges is obscure in the literature . The main ingredients of the current theory are the  existence of two symmetries which are connected by hierarchy and the emergence of tensor gauge model . In this letter we will derive all these as logical consequences  of our model.
       
         \hskip 1 cm         The root cause of the (near) zero  mobility of the electrons was discovered \cite{nand}. A theory was advanced which was sufficient to explain most of the phenomenology but   a pure gauge theory was lacking. However we will see that the dynamics of the  charged matter which was so far not adequately treated,is the most crucial thing .  If one chooses the complex klein Gordon field to determine the dynamics of the charged matter then there is no scope for the 2nd gauge   symmetry.   We  show in this letter that it is possible to build fracton like symmetries from a theory involving the complex scalar field. The point of departure is  {\bf{the basic building blocks of the theory are no longer}} the zero mass Klein Gordon Lagrangian, a lesser known but a deep theory -- it is the Galileon scalar field {\cite{nic}}.



(ii){\bf{Introduction of  the Galileon based  charged scalar fields :}} 
 
                   The first task is to define the matter fields representing the fracton system. As discussed in introduction the  Galileon model will be the building block of the charged matter theory. Thus we define 
\begin{eqnarray}
\phi &=&\pi_{1}+i\pi_{2} \nonumber\\
\phi^{*}&=&\pi_{1}-i\pi_{2} \label{98}
\end{eqnarray} 
 where $\pi_{1}$,$\pi_{2}$ are Galileon fields , including upto ${\mathcal{L}}_{3}$, where
 
 \begin{eqnarray}
 {\mathcal{L}}_{3}=\Box\pi\partial_{\mu}\pi\partial^{\mu}\pi
 \end{eqnarray}
 
 We require to construct the Lagrangian of our theory in terms of $\phi$ and $\phi^{*}$. The Lagrangian must satisfy certain requirements
 
a.
It must be real.

b.
It must be Poincare invariant.

c.
It must be symmetric under global phase rotation.

d.
It will have  shift symmetry with the parameters extended to complex values.

e.
The equation of motion must be second order in time derivative of $\phi$ and $\phi^{*}$.

 Hence we propose the Lagrangian, 

\begin{eqnarray}
 {\mathcal{L}}= \bigg(\phi\Box{\phi^{*}}+\phi^{*}\Box{\phi}\bigg)\partial_{\mu}\phi\partial^{\mu}\phi^{*}
 \label{1}
\end{eqnarray}
 
The justification of 
(\ref{1}) can be established by observing that  it is symmetric in $\phi $ and
 $\phi^*$ , invariant under a kind of shift symmetry with complex parameter ,invariant under spacetime transformations, invariant under global phase rotation  etc .  The proof of the statement will now be given  below.

(iii){ \bf{ Equation of motion : }} 

The specific Lagrangian (\ref{1}) which is a function  $\phi$ and  $\phi^{*}$, of the fields and there derivatives upto the second order . Note that one will expect
that the equations of motion would contain time derivatives of the fields of third order. But the Galileon model is so constructed that such terms will cancel out.
\begin{eqnarray} 
 S= \int{\mathcal{L}}(\phi,\partial_{\mu}\phi,\partial_{\mu}\partial^{\nu}\phi)d^{4}x \nonumber
\end{eqnarray}

 As usual, let there be a variation $\delta\phi$ which vanishes at the initial and final instants as well as the spatial boundaries (which is usually at infinite). Hence the equation of motion are,

\begin{eqnarray}
 \frac{\delta S}{\delta \phi}= \bigg[\frac{\partial{\mathcal{L}}}{\partial\phi} -\partial_{\mu}\bigg(\frac{\partial{\mathcal{L}}}{\partial(\partial_{\mu}\phi)}\bigg)+\partial_{\mu}\partial^{\nu}\bigg(\frac{\partial{\mathcal{L}}}{\partial(\partial_{\mu}\partial^{\nu}\phi)}\bigg)\bigg] =0\label{4}
\end{eqnarray}

Using the  Lagrangian ({\ref{1}}) and the theorem that $(\partial\partial\pi)^{n}$ =0 \cite{nic}, we get the equation of motion of our model as
\begin{eqnarray}
\Box{\phi^{*}}(\partial_\mu \phi \partial^{\mu} \phi^* - \phi\Box{\phi^{*}} )- \Box{\phi}(\partial_\mu \phi^{*} \partial^{\mu} \phi^* - \phi^{*}\Box{\phi^{*}})  =0
\label{eom}
\end{eqnarray} 
We can easily realize that the only solutions to (\ref{eom}) is
\begin{eqnarray}
\Box{\phi^{*}}= \Box{\phi}=0
\label{eom1}
\end{eqnarray} 
 
 Similarly the equation of motion for $\phi^{*}$ may be obtained. In our theory the independent fields are $\phi$ and $\phi^{*}$. Where $\phi^{*}$ is the complex conjugate of $\phi$.The most remarkable aspects of the equations of motion is that they contain no  derivative of higher order than the second. Note that this result is standard for the Galileon model. Thus our construction of the charged scalar field does not contradict this property of Galileon .This is very important because the Lagrangian ({\ref{1}}) 
 depends on the  higher derivative but is devoid of ostrogradsky ghosts \cite{ostro}.  Thus the theory proposed here can be considered as the complex continuation of Galileon theory \cite{nic}. As far as we know this is the first occurrence of  such extension of Galileon model in the literature. 
 
(iv){ \bf{ Proof of different symmetries : }} 
 
 As stated about our model (\ref{1}) is endowed with various symmetries.     The ${\mathcal{U}}(1)$ symmetry is obvious. If we substitute $\phi \to e^{i\alpha}\phi $ in (\ref{1}) then we see that the Lagrangian is unchanged. Carefully note that this will only be true when the parameter $\alpha$ is constant (i.e only for global phase rotation). One also should note that the above invariance is directly a symmetry  of the action that is off shell . The charged matter interact electromagnetically which has the  ${\mathcal{U}}(1)$ symmetry group. This is a continuous  Abelian group with one generator. Naturally it cannot have any more off shell invariance but then how do we claim that our Galileon based action has two different symmetries ? This is possible for the Galileon, which has a shift symmetry (\ref{0}) belongs to the  symmetry group of the equation of motion \cite{nic}  \footnote{For instance consider their comment\cite{nic} ``As follows from our discussion we are thus lead to make a second non trivial assumption, namely that the equations of motion of $\pi$ be invariant under shift symmetry (\ref{0})'' . Thus we see that the status of the second symmetry of  the Galileon based model is on shell. }. 

Slightly more elaborate is the space time symmetry. For space time transformation the coordinates transforms as   
\begin{eqnarray}
x \to x +\Lambda x
\end{eqnarray}

 where $\Lambda$ is the Lorentz matrix. Then according to index theorem in tensor analysis  all the terms of Lagrangian (\ref{1}) are separately scalar. Hence the theory is invariant under Lorentz transformations. Similarly it can be shown that the theory has transnational invariance\footnote{However the transnational invariance is on shell.}. Hence our theory is symmetric under the full Poincare group.

Now the most nontrivial question  is whether the theory (\ref{1}) has shift symmetry or not. Here the Galileon scalars $\pi_1$
and $\pi_2$ transform as,

\begin{eqnarray} 
 \pi_{1} \to \pi_{1}+ a+b_{\mu}x^{\mu}\nonumber\\
 \pi_{2} \to \pi_{2}+ a+b_{\mu}x^{\mu} \label{12}
  \end{eqnarray}
  
   Using the defining relations  we obtain

\begin{eqnarray}
\phi &\to& \bigg(\phi +(1+ i) (a+b_{\mu}x^{\mu})\bigg) \nonumber\\
\phi^{*} &\to &\bigg(\phi^{*} + (1 - i) (a+ b_{\mu}x^{\mu})\bigg)\label{mo}
\end{eqnarray} 

Thus for shift transformation (\ref{mo}) we see , $\delta\phi= (1+i) (a+ b_{\mu}x^{\mu})$ , $\delta\phi^{*}=(1-i)(a + b_{\mu}x^{\mu})$,  $\delta(\partial_{\mu}\phi)=(1+i)b_{\mu}$  ,  $\delta(\partial_{\mu}\phi^{*})=(1-i)b_{\mu}$ , $\delta(\Box{\phi})=0$ and $\delta(\Box{\phi^{*}})=0$ . 

 The variation of (\ref{1}) due to shift transformation is given by
 
  \begin{eqnarray}
 {\delta\mathcal{L}}&=&\bigg[\delta(\phi) \Box{\phi^{*}}+\delta(\phi^{*})\Box{\phi}+\phi \delta(\Box{\phi^{*}})+\phi^{*} \delta(\Box{\phi})]\partial_{\mu}\phi\partial^{\mu}\phi^{*}\nonumber\\          &  &\qquad \qquad  + (\phi \Box{\phi^{*}}+\phi^{*}\Box{\phi})  \bigg[\delta(\partial_{\mu}\phi)\partial^{\mu}\phi^{*} + \partial_{\mu}\phi\delta(\partial^{\mu}\phi^{*})\bigg]
\end{eqnarray}
As we have discuss at the beginning of the section this symmetry is on shell, so using equation of motion we get
  
   \begin{eqnarray}
 {\delta\mathcal{L}}= 0
\end{eqnarray}
 Thus our proposed complex scalar model (\ref{1}) should have a new type of shift symmetry namely the analytic continuation of the Galileonic shift symmetry.


Point should be noted here that the phase rotation symmetry is global gauge symmetry of the action where as the complex shift symmetry (\ref{mo}) is holds modulo the equation of motion. This fact , which has been relieved from our research explains ,
\begin{enumerate}
\item
The reason of , on the experimental side , there exist only  one kind of mediator (photon) rather than  two .
\item
On the theoretical side , the existence of two different symmetries while the gauge space is one dimensional.
\end{enumerate}

\hspace{1 cm}             Another issue is the number of gauge symmetries. One may ask are these two symmetries
independent? To answer this question note that the  Lagrangian (\ref{1}) is a higher derivative theory. For the usual theories
(like the Klein Gordon theory) the Lagrangian contains at most second order derivatives of fields. It is known for such  theory that the number of independent gauge degrees of freedom  equals the number of primary first class constraints in the phase space \cite{dec}. Converting the higher derivative theory to an equivalent first order theory \cite{w} it was proved that in higher derivative theory the number of gauge degrees of freedom may be less than the number of primary first class constraints where the first class constraints corresponds to gauge symmetry. In our case the physical interaction is the electromagnetic interaction which is responsible for both the symmetry . So there  will be some correspondence between the shift transformation and the phase rotation transformation. We will use this point in appropriate place.

 

\item{{\bf{ Conservation laws:}}}

     We have seen that Galileon charged matter  theory  has two additional  gauge-like symmetries, the first of these is the well known phase rotation symmetry and the other one is analytic continuation of the shift symmetry.   We must see what conservation laws are indicated by it. A useful  procedure is to apply Noether's theorems. Due to  higher derivatives in the Lagrangian the Noether's formula will be appropriately modified. So  the conservation law is better to be derived from the functional derivative with respect to the parameters of transformation. First let us derive a general formula.  This can then be specialized to different symmetry transformations of (\ref{1}).


          If $\delta\phi$ be the change in $\phi$ then the action  changes by,
\begin{eqnarray}
  \delta S &= & \int\bigg[\frac{\partial{\mathcal{L}}}{\partial\phi}\delta \phi+\frac{\partial{\mathcal{L}}}{\partial\phi^{*}}\delta \phi^{*} +\frac{\partial{\mathcal{L}}}{\partial(\partial_{\mu}\phi)}\delta(\partial_{\mu} \phi)+\frac{\partial{\mathcal{L}}}{\partial(\partial_{\mu}\phi^{*})}\delta(\partial_{\mu} \phi^{*})     \nonumber\\          &  &\qquad \qquad   +\frac{\partial{\mathcal{L}}}{\partial(\partial_{\mu}\partial^{\nu}\phi)}\delta(\partial_{\mu}\partial^{\nu} \phi)+\frac{\partial{\mathcal{L}}}{\partial(\partial_{\mu}\partial^{\nu}\phi^{*})}\delta(\partial_{\mu}\partial^{\nu} \phi^{*})\bigg]d^{4}x \nonumber
   \label {ma}
\end{eqnarray}
Note that $\delta\phi$ not necessarily vanish on the boundary. Using equations of motion we obtain

\begin{eqnarray}
\delta S &=& \int\partial_{\sigma}\bigg[ \frac{\partial{\mathcal{L}}}{\partial(\partial_{\sigma}\phi)}\delta\phi  +\frac{\partial{\mathcal{L}}}{\partial(\partial_{\sigma}\phi^{*})}\delta\phi^{*}  -\partial^{\lambda}\bigg(\frac{\partial{\mathcal{L}}}{\partial(\partial_{\sigma}\partial^{\lambda}\phi)}\bigg)\delta \phi  \nonumber\\          &  &\qquad \qquad -   \partial^{\lambda}\bigg(\frac{\partial{\mathcal{L}}}{\partial(\partial_{\sigma}\partial^{\lambda}\phi^{*})}\bigg)\delta \phi^{*} \bigg] d^{4}x \label{211}
\end{eqnarray}

Equation (\ref{211}) is the master formula for calculating the conserved currents and charges. Now, we will specialize for different symmetries. Start with the shift symmetry . For shift transformation (\ref{mo}) we see , $\delta\phi= (1+ i) (a+ b_{\mu}x^{\mu})$ , $\delta\phi^{*}=(1-i)(a + b_{\mu}x^{\mu})$, substituting this variations of $\phi$ and $ \phi^*$ in (\ref{211}) and after some algebra
we obtain expressions 

\begin{eqnarray}
{\theta}^{\sigma}= -(a+ b_{\mu}x^{\mu})\bigg[(1+i)\partial^{\sigma}\phi^{*}\partial_{\alpha}\phi\partial^{\alpha}\phi^{*}+(1-i)\partial^{\sigma}\phi\partial_{\alpha}\phi\partial^{\alpha}\phi^{*}  \bigg]\label{13}
\end{eqnarray}
 thus the current corresponding to $b_\mu$
is given by

\begin{eqnarray}
{\theta^{\sigma\mu}}_{b}= -x^{\mu}\bigg[(1+i)\partial^{\sigma}\phi^{*}\partial_{\alpha}\phi\partial^{\alpha}\phi^{*}+(1-i)\partial^{\sigma}\phi\partial_{\alpha}\phi\partial^{\alpha}\phi^{*}\bigg]\label{17}
\end{eqnarray}
Note that the current is a two indexed quantity . So the corresponding charge is 

\begin{eqnarray}
{J^k}_b =-\int {\theta^{0 k}}_{b}d^3x = -\int x^{k}\bigg[(1+i)\dot{\phi^{*}}\partial_{\alpha}\phi\partial^{\alpha}\phi^{*}+(1-i)\dot{\phi}\partial_{\alpha}\phi\partial^{\alpha}\phi^{*} \bigg]d^3x
\label{171}
\end{eqnarray}

Again the conserved current corresponding to $a$ is given by

\begin{eqnarray}
{\theta^{\sigma}}_{a}=-\bigg[(1+i)\partial^{\sigma}\phi^{*}\partial_{\alpha}\phi\partial^{\alpha}\phi^{*}+(1-i)\partial^{\sigma}\phi\partial_{\alpha}\phi\partial^{\alpha}\phi^{*}  \bigg]\label{18}
\end{eqnarray}

So the corresponding charge is 

\begin{eqnarray}
\int {\theta^{0}}_{a}d^3x = -\int\bigg[(1+i)\dot{\phi^{*}}\partial_{\alpha}\phi\partial^{\alpha}\phi^{*}+(1-i)\dot{\phi}\partial_{\alpha}\phi\partial^{\alpha}\phi^{*}  \bigg]d^3x
\label{172}
\end{eqnarray}

In the following we will argue that both the Galileonic symmetries (following from $a$ and $b_{\mu}$) are consequence of electromagnetic interaction. This observation has a crucial impact as  shall see.

From equations (\ref{171}) and (\ref{172})  we can establish that the conserved `charge density' is 
\begin{eqnarray}
\rho_a =  -\bigg[(1+i)\dot{\phi^{*}}\partial_{\alpha}\phi\partial^{\alpha}\phi^{*}+(1-i)\dot{\phi}\partial_{\alpha}\phi\partial^{\alpha}\phi^{*} \bigg]
\end{eqnarray}

It is now remarkable to observe that the conserved charge corresponding to shift is the `dipole moment' corresponding to the charge density $\rho_a $  i.e.

\begin{equation}
{J^k}_b = \int \rho_a(x)x^k d^3 x \label{gobbor}
\end{equation} 

At this point we should pause and ask: why  is   the charge? To understand that we may remember that we are considering an `` effective charge" , which the Galileonic excitations offer to the applied field . But this charge is also determined by the phase rotation symmetries as ,

\begin{eqnarray}
\int \theta^{0}d^3x = -i \int \bigg[\bigg(\phi\dot{\phi^{*}}\partial_{\alpha}\phi\partial^{\alpha}\phi^{*}-\phi^{*}\dot{\phi}\partial_{\alpha}\phi\partial^{\alpha}\phi^{*}  \bigg)\bigg]d^3x
\label{188}
\end{eqnarray}

From equation (\ref{172}) and (\ref{188}) we can find a correspondence of the charges between the shift symmetry and the phase rotation parameter  \cite{remark}. This correspondence is possible only because the fracton matter action is a higher derivative theory \cite{w}. As a consequence (\ref{gobbor}) signifies a `electric dipole moment'. Note that the conservation of dipole moment can only hold in a coordinate invariant  manner subject to the conservation of charge . The two symmetries therefore are not independent. So the riddle that so far remain a mystery  in the existing fracton gauge theory \cite{pret1} \cite{nand}  follow smoothly in our result.

\item{\bf{Interaction of the fractons :}}

 We have identified three different symmetries of theory (\ref{1}) namely the space time symmetry, the phase rotation symmetry and the complex  shift symmetry\footnote{Here transformation parameter are complex, originally the term shift symmetry was used in Galileon field. }. We know  that the phase rotation symmetry in the gauge space produces an electromagnetic interaction.  So fractons will be able to interact electromagnetically . Note that the complex shift symmetry does not produce a different mediator but constrains the form of elementary excitation in the form of dipoles. So we can explain analytically the dipole interaction mooted phenomenologically earlier \cite{pret}.

\hskip 1 cm For understanding the dynamics of the fracton and the electromagnetic field we have to localize the shift  symmetry and phase rotation symmetry by making the  parameters ($\alpha$ , $b_{\mu}$) arbitrary function of space and time \footnote{ As there is some correspondence between the shift symmetry parameter $ a $   and the phase rotation parameter $\alpha$ , so we does not need to localize $ a $ \cite{remark}.}. Now  localization  parameter $ \alpha $  results in a vector gauge theory , where as localization of $b_{\mu}$ will  introduced a new covariant quantity which must be tensor gauge field. Thus we have to replace $\partial_\mu $ by 

\begin{eqnarray}
 D_{\mu} {\phi}= \partial_{\mu}\phi +iA_{\mu}\phi+(1+i)B_{\mu\nu} x^{\nu}\phi \label{pk}
 \end{eqnarray} 
 Where $A_{\mu}$ and $ B_{\mu\nu}$ are the new fields which respectively are a vector and a symmetric tensor. The transformation of $A_{\mu}$ and $ B_{\mu\nu}$ will ensure that the``new derivative "($D_{\mu}\phi$) will transform in the same way as the usual derivatives do in the global theory. A complete analysis of this problem is out of the scope of this letter . But it appears that this tensor field $ B_{\mu\nu}$ will have similar role as does the symmetric tensor gauge field in the phenomenology of fractons {\cite{pret1}}.
   
\hskip .5cm   Last but not the least, we have said nothing about the space time symmetries of the fracton. For Galileons we know that the gauging of the space time symmetries lead to an interaction which is very similar to modified gravity \cite{geon}\cite{pm}(that is Galileons have a gravitational interaction in the curved space time background). As fracton action is produced by Galileon doublet , it is quite probable that fracton will also have a gravitational interaction in the curved  background . Note that it was earlier conjectured  \cite{pret2} that the symmetric tensor gauge field which belongs to the generalized Maxwell's theory may be responsible for gravitational interaction also. But from our analysis it is quite clear that symmetric tensor gauge field $B_{\mu\nu}$ associated with shift symmetry (see equation \ref{pk}) has nothing to do with the emergent gravity.

\item{\bf{ Discussion of the results and the concluding remarks :
}}

It is time now to consolidate what we have achieved. We assume that matter in fracton phase is described by a charged scalar field , where the basic elements are Galileon scalars . Galileons are  related  with Horndeski's  analysis of higher order Lagrangian theories. From this one can show that the equation of motion that follow is of usual  second order in time.   From the results of model the Galileonic nature of the fracton phases of matter    is  justified. Note that our theory has the full Lorentz symmetry and translation symmetry. So at the covariant level the space time symmetries are apparent. This conclusion is significant in the back drop of the results reported in the literature.

\hskip    1 cm  Thus we find that the research reported in this letter leads to a paradigm shift in the understanding of the fracton phase of matter . Since the  fracton phenomena is reported to have far reaching applications  in such vast area as  fracton-elasticity duality , three dimensional crystals , glassy dynamics , new condensed matter platform as Majorana Islands , Hole-Doped AntiFerromagnetics etc \cite{pret1} , our discovery will have a prominent role in the future analysis.

\end{enumerate}

{\bf{Aknowledgments}}

 Sk. Moinuddin would like to thank CSIR India for the fellowship provided to him.(File no: 08/606(0005)/2019-EMR-I)  and 
P M thanks Rabin Banerjee for introducing him to \cite{pret}.


\begin{thebibliography}{999}
\bibitem{remark}
Similar correspondence has been assumed in \cite{and}.
\end{thebibliography}
\end{document}